\documentstyle[12pt]{article}

\font\twlgot =eufm10 scaled \magstep1
\font\egtgot =eufm8
\font\sevgot =eufm7

\font\twlmsb =msbm10 scaled \magstep1
\font\egtmsb =msbm8
\font\sevmsb =msbm7

\newfam\gotfam

\textfont\gotfam\twlgot
\scriptfont\gotfam\egtgot
\scriptscriptfont\gotfam\sevgot

\newfam\msbfam
\textfont\msbfam\twlmsb
\scriptfont\msbfam\egtmsb
\scriptscriptfont\msbfam\sevmsb
\def\Bbb{\protect\pBbb}
\def\pBbb{\relax\ifmmode\expandafter\Bb\else\typeout{You cann't use
Bbb in text mode}\fi}
\def\Bb #1{{\fam\msbfam\relax#1}}

\def\thebibliography#1{\bigskip\section*{\centering
References\\}\bigskip\list
  {\arabic{enumi}.}{\settowidth\labelwidth{#1}\leftmargin\labelwidth
    \advance\leftmargin\labelsep
    \usecounter{enumi}}
    \def\newblock{\hskip .11em plus .33em minus .07em}
    \sloppy\clubpenalty4000\widowpenalty4000
    \sfcode`\.=1000\relax}

\newcommand{\Si}{\Sigma}

\def\op#1{\mathop{\fam0 #1}\limits}

\def\Ker{{\rm Ker\,}}

\newcommand{\ben}{\begin{eqnarray}}
\newcommand{\een}{\end{eqnarray}}
\newcommand{\be}{\begin{eqnarray*}}
\newcommand{\ee}{\end{eqnarray*}}
\newcommand{\bea}{\begin{eqalph}}
\newcommand{\eea}{\end{eqalph}}

\newcommand{\cL}{{\cal L}}
\newcommand{\cE}{{\cal E}}
\newcommand{\cH}{{\cal H}}
\newcommand{\cF}{{\cal F}}

\newcommand{\al}{\alpha}
\newcommand{\bt}{\beta}

\newcommand{\la}{\lambda}
\newcommand{\La}{\Lambda}

\newcommand{\p}{\pi}
\newcommand{\s}{\psi}

\newcommand{\om}{\omega}
\newcommand{\Om}{\Omega}
\newcommand{\m}{\mu}

\newcommand{\g}{\gamma}
\newcommand{\G}{\Gamma}
\newcommand{\e}{\epsilon}
\newcommand{\ve}{\varepsilon}
\newcommand{\th}{\theta}

\newcommand{\si}{\sigma}

\newcommand{\w}{\wedge}

\newcommand{\wt}{\widetilde}
\newcommand{\wh}{\widehat}
\newcommand{\ol}{\overline}
\newcommand{\dr}{\partial}

\newcounter{eqalph}
\newcounter{equationa}

\newenvironment{eqalph}{\stepcounter{equation}
\setcounter{equationa}{\value{equation}}
\setcounter{equation}{0}

\begin{eqnarray}}{\end{eqnarray}
\setcounter{equation}{\value{equationa}}}

\begin{document}
\hbox{}

\centerline{\bf\large GRAVITY AS A HIGGS FIELD.}
\medskip

\centerline{\bf\large II. FERMION-GRAVITATION COMPLEX}
\bigskip

\centerline{\bf Gennadi A Sardanashvily}
\medskip

\centerline{Department of Theoretical Physics, Physics Faculty,}

\centerline{Moscow State University, 117234 Moscow, Russia}

\centerline{E-mail: sard@theor.phys.msu.su}
\bigskip

\centerline{\bf Abstract}
\medskip

{\it If gravity is a metric field by Einstein, it is a Higgs field.}
Gravitation theory meets spontaneous symmetry breaking when the
structure group of the principal linear frame bundle $LX$ over a world
manifold $X^4$ is reducible to the connected Lorentz group $SO(3,1)$.
The physical underlying reason of this reduction is Dirac fermion matter
possessing only exact Lorentz symmetries. The associated Higgs field is
a tetrad gravitational field $h$ represented by a global section of the
quotient $\Si$ of $LX$ by $SO(3,1)$. The feature of gravity as a Higgs
field issues from the fact that, in the presence of different tetrad fields,
there are nonequivalent representations of cotangent vectors to $X^4$ by
Dirac's matrices. It follows that, in gravitation theory, fermion fields
must be regarded only in a pair with a certain tetrad field. These pairs
constitute the so-called fermion-gravitation complex and are represented
by sections of the composite spinor bundle $S\to\Si\to X^4$ where values
of tetrad gravitational fields play the role of coordinate parameters,
besides the familiar world coordinates. In Part 1 \cite{sard10} of the
work, geometry of this composite spinor bundle has been investigated.
This Part is devoted to dynamics of fermion-gravitation complex. It is
a constraint system to describe which we use the covariant multisymplectic
generalization of the Hamiltonian formalism when canonical momenta
correspond to derivatives of fields with respect to all world coordinates,
not only the time. On the constraint space, the canonical momenta of a
tetrad gravitational field as a Higgs field are equal to zero, otherwise
in the presence of fermion fields. Fermion fields deform the constraint
space in the gravitation sector that leads to modification of the Einstein
equations.

\section{Introduction}

Gravitation theory is theory with spontaneous symmetry breaking
established by the equivalence principle reformulated
in the terms of Klein-Chern geometries of invariants \cite{iva,tsar,3sar}.
It postulates that there exist reference
frames with respect to wich Lorentz invariants can be defined everywhere
on a world manifold $X^4$. This principle has
the adequate mathematical formulation in terms of fibre bundles.

Let $LX$ be the principal bundle of linear frames in tangent spaces to
$X^4$.
The geometric equivalence principle requires that its structure group
\[
GL_4=GL^+(4,{\bf R})
\]
is reduced to the connected Lorentz group
\[
L=SO(3,1).
\]
It means that there is given a reduced subbundle $L^hX$ of $LX$ whose
structure group is $L$. They are atlases of $L^hX$ with respect to which
Lorentz invariants can be defined.
In accordance with the well-known theorem, there is
the 1:1 correspondence between the reduced $L$ subbundles $L^hX$ of
$LX$ and the tetrad gravitational fields $h$ represented by global
sections of the Higgs bundle
\begin{equation}
\Si=LX/L\to X^4 \label{5.15}
\end{equation}
with standard fibre $GL_4/L$.

The underlying physical reason of the geometric equivalence principle
is Dirac fermion matter possesing only exact Lorentz symmetries.

Let us consider a bundle of complex Clifford algebras ${\bf C}_{3,1}$
over $X^4$. Its subbundles are both a spinor bundle $S_M\to X^4$ and the
bundle $Y_M\to X^4$ of Minkowski spaces of generating elements of
${\bf C}_{3,1}$. There is the bundle morphism
\[
\g: Y_M\otimes S_M\to S_M
\]
which defines representation of elements of $Y_M$ by Dirac's
$\g$-matrices on elements of the spinor bundle $S_M$. To describe Dirac
fermion fields on a world manifold, one must require that the bundle
$Y_M$ is isomorphic to the cotangent bundle $T^*X$ of $X^4$. It takes
place if $Y_M$ is associated with some reduced $L$ subbundle $L^hX$
of the linear frame bundle $LX$. Then, there exists the representation
\[
\g_h : T^*X\otimes S_h\to S_h
\]
of cotangent vectors to a world manifold $X^4$ by Dirac's $\g$-matrices
on elements of the spinor bundle $S_h$ associated with the lift of $L^hX$
to a $SL(2,{\bf C})$ principal bundle. Sections of $S_h$ describe
Dirac fermion fields in the presence of a tetrad gravitational field $h$.

The crusial point consists in the fact that,
for different tetrad fields $h$ and $h'$, the representations $\g_h$ and
$\g_{h'}$ fail to be equivalent. It follows that every
Dirac fermion field must be regarded only in a pair with a certain
tetrad gravitational field $h$. There is the 1:1 correspondence between
these pairs and the sections of the composite bundle
\begin{equation}
S\to\Si\to X^4 \label{L1}
\end{equation}
where $S\to\Si$ is a spinor bundle associated with the $L$ principal
bundle $LX\to\Si$ \cite{3sar,sard10}.

This Part of the work covers dynamics of the fermion-gravitation complex.

Dynamics of fields represented by sections of a fibred
manifold $Y\to X$ is phrased in terms of jet manifolds
\cite{gia,got,kup,sard}.

Recall that the $k$-order jet manifold $J^kY$ of a fibred
manifold $Y\to X$ comprises the equivalence classes
$j^k_xs$, $x\in X$, of sections $s$ of $Y$ identified by the $(k+1)$
terms of their Taylor series at $x$. It is a finite-dimensional manifold.
Jet manifolds have been widely used in the
theory of differential operators. Their application to differential
geometry is based on the  1:1 correspondence between the connections
on a fibred manifold $Y\to X$ and the global sections of the jet bundle
$J^1Y\to Y$.

In the first order Lagrangian formalism,
the jet manifold $J^1Y$ plays the role of a finite-dimensional
configuration space of fields. Given fibred coordinates $(x^\la,y^i)$
of $Y\to X$, it is endowed with the adapted
coordinates $(x^\la,y^i,y^i_\la)$ where coordinates $y^i_\la$ make the
sense of values of partial derivatives $\dr_\la y^i(x)$ of field
functions $y^i(x)$. A Lagrangian density on $J^1Y$ is defined by a form
\[
L=\cL(x^\la,y^i,y^i_\la)\om, \qquad \om=dx^1\w...\w dx^n, \qquad n=\dim X.
\]
If a Lagrangian density is degenerate, the corresponding
Euler-Lagrange equations are underdetermined and need supplementary
gauge-type conditions. In gauge theory, they are the familiar gauge
conditions. In general case, the above-mentioned supplementary
conditions remain elusive.

To describe constraint field systems, one
can use the covariant multimomentum Hamiltonian formalism where
canonical momenta correspond to derivatives of fields with respect
to all world coordinates, not only the time
\cite{car,gun,6sar,sard,lsar}. Given a fibred manifold $Y\to X$, the
corresponding multimomentum phase space is the Legendre bundle
\begin{equation}
\Pi=\op\w^n T^*X\op\otimes_Y TX\op\otimes_Y V^*Y \label{00}
\end{equation}
over $Y$ into which the Legendre morphisms $\wh L$ associated with
Lagrangian densities $L$ on $J^1Y$ take their values.
This bundle is provided with the fibred coordinates $(x^\la ,y^i,p^\la_i)$
such that
\[
 (x^\m,y^i,p^\m_i)\circ\wh L=(x^\m,y^i,\pi^\m_i),
\qquad \pi^\mu_i=\dr^\mu_i\cL.
\]
We shall call them the canonical coordinates.
The Legendre bundle (\ref{00}) carries the generalized Liouville form
\begin{equation}
 \th =-p^\la_idy^i\w\om\otimes\dr_\la	\label{2.4}
\end{equation}
and the multisymplectic form
\begin{equation}
\Om =dp^\la_i\w
dy^i\w\om\otimes\dr_\la. \label{406}
\end{equation}
If $X=\Bbb R$, they recover respectively the
Liouville form and the symplectic form in mechanics.

The multimomentum Hamiltonian formalism is phrased intrinsically in terms
of Hamiltonian connections which
play the  role similar Hamiltonian vector fields in the symplectic geometry.
We  say that a connection
$\g$ on the fibred Legendre manifold $\Pi\to X$ is a Hamiltonian
connection if the  form  $\g\rfloor\Om$ is closed.
Then, a Hamiltonian form $H$ on $\Pi$ is defined to be an
exterior form such that
\begin{equation}
dH=\g\rfloor\Om \label{013}
\end{equation}
for some Hamiltonian connection $\g$. The key point consists in the
fact that every Hamiltonian form admits splitting
\begin{equation}
H =p^\la_idy^i\w\om_\la
-p^\la_i\G^i_\la\om
-\wt{\cH}_\G\om=p^\la_idy^i\w\om_\la-\cH\om \label{017}
\end{equation}
where $\G$ is a connection on the fibred manifold $Y$ and
\[
\om_\la=\dr_\la\rfloor\om.
\]
Given the  Hamiltonian form $H$ (\ref{017}), the equality
(\ref{013}) comes to the Hamilton equations
\bea
&&\dr_\la y^i(x) =\dr^i_\la\cH, \label{3.11a}\\
&& \dr_\la p^\la_i(x) =-\dr_i\cH \label{3.11b}
\eea
for sections $r$ of the fibred Legendre manifold $\Pi\to X$.

If a Lagrangian density $L$ is regular, there exists the unique Hamiltonian
form $H$ such that the first order Euler-Lagrange equations and the
Hamilton equations are equivalent, otherwise in case of degenerate
Lagrangian densities.
Given a degenerate Lagrangian system when the constraint
space is $Q=\wh L(J^1Y)$, one must consider a
family of different  Hamiltonian forms $H$ associated with the same
Lagrangian density $L$ in order to exaust solutions of the Euler-Lagrange
equations.

Lagrangian densities of field models are almost always quadratic and
affine in derivative coordinates $y^i_\mu$.
In this case, we have the comprehensive relation between
solutions of the Euler-Lagrange equations and the Hamilton equations.
Given an associated Hamiltonian form $H$, every solution of the
corresponding Hamilton equations which
lives on the constraint space $Q$ yields
a solution of the Euler-Lagrange equations. Conversely,
for any solution of the Euler-Lagrange equations, there
exists the corresponding solution of the Hamilton equations for some
associated Hamiltonian form.
In particular, Hamilton equations are
separated in the dynamic equations and the above-mentioned gauge-type
conditions independent of momenta $p^\la_i$.

It is the multimomentum Hamiltonian formalism which enables one to
analyse dynamics of field systems on composite manifolds, in particular,
dynamics of fermion-gravitation complex.

In the gauge gravitation theory, classical gravity is described by pairs
$(h,A_h)$ of tetrad gravitational fields $h$ and gauge gravitational
potentials $A_h$ identified with principal connections on the reduced
$L$ subbundles $L^hX$ of the linear frame bundle $LX$.
Every connection on $L^hX$ is extended
to a Lorentz connection on $LX$ which however fails to be reducible
to a principal connection on another reduced subbundle $L^{h'}X$ if
$h\neq h'$. It follows that gauge gravitational potentials also must be
regarded in pairs with a certain tetrad gravitational field $h$.
Following the general procedure \cite{sard,sard10}, one can describe these
pairs $(h,A_h)$ by sections of the composite bundle
\begin{equation}
C_L=J^1LX/L\to J^1\Si\to\Si\to X^4. \label{N53}
\end{equation}
It is endowed with the local fibred coordinates
\[
(x^\mu,\si^\mu_a,k^{ab}{}_\la=
-k^{ba}{}_\la, \si^\mu_{a\la})
\]
where $(x^\mu,\si^\mu_a, \si^\mu_{a\la})$ are coordinates
of the jet bundle $J^1\Si$. Given a section $s$ of $C_L$,
we recover familiar tetrad functions and Lorentz gauge potentials
\[
(\si^\mu_a\circ s)(x)=h^\mu_a(x), \qquad (k^{ab}{}_\la \circ
s)(x)=A^{ab}{}_\la(x)
\]
respectively.
The corresponding configuration space is the jet manifold $J^1C_L$.

The total configuration space of the fermion-gravitation complex is
the product
\begin{equation}
J^1C_L\op\times_{J^1\Si}J^1S \label{M3}
\end{equation}
where $S$ is the composite spinor bundle (\ref{L1}) endowed with the
coordinates $(x^\mu,\si^\mu_a,y^A)$.

As a test case, we shall restrict our consideration to the
Hilbert-Einstein Lagrangian density. Then, on the configuration space
(\ref{M3}), the feature of the fermion-gravitation complex lies only in the
generalization
\[
\wt D_\la =y^A_\la- \frac12(k^{ab}{}_\la +N^{ab}{}_\la)
I_{ab}{}^A{}_By^B
\]
of the familiar covariant differential of Dirac fermion fields by means of
the composite term
\[
N^{ab}{}_\la= A^{ab}{}^c_\mu (\si^\mu_{c\la} -\G^\mu_{c\la}).
\]
Here, $\G^\mu_{c\la}$ is a connection on the Higgs bundle $\Si$ and
\begin{equation}
A^{ab}{}^c_\mu=\frac12(\eta^{ca}\si^b_\mu -\eta^{cb}\si^a_\mu)
\label{M4}
\end{equation}
where $\eta$ denotes the Minkowski metric is the
canonical connection on the bundle
\begin{equation}
GL_4\to GL_4/L. \label{M5}
\end{equation}
As an immediate consequence, the standard gravitational constraints
\begin{equation}
p^{c\la}_\mu=0 \label{M1}
\end{equation}
where $p^{c\la}_\mu$ are canonical momenta of tetrad fields are replaced
by the relations
\begin{equation}
p^{c\la}_\mu+\frac12A^{ab}{}^c_\mu I_{ab}[]^A{}_By^Bp^\la_A=0 \label{M2}
\end{equation}
where $p^\la_A$ are canonical momenta of fermion fields. When restricted
to the constraint space (\ref{M1}), the Hamilton equations (\ref{3.11b})
\[
\dr_\la p^{c\la}_\mu =\dr^c_\mu\cH
\]
come to the familiar Einstein equations, otherwise on the constraint space
(\ref{M2}).

Thus, the goal is modification of the
Einstein equations for the total system
of fermion fields and gravity because of deformation of gravitational
constraints.
This deformation makes also contribution into the energy-momentum
conservation law. In the framework of the multimomentum Hamiltonian
formalism, we have the fundamental identity whose restriction to a
constraint space can be treated as the energy-momentum conservation law
\cite{sard}. In Part III of the work, percularity of
this conservation law in gravitation theory will be considered.

\section{Technical preliminary}

Given a fibred manifold $Y\to X$, the first order jet manifold $J^1Y$ of
$Y$ is both the fibred manifold $J^1Y\to X$
and the affine bundle $J^1Y\to Y $  modelled on the vector
bundle  $T^*X\otimes_Y VY. $
The adapted coordinates $(x^\la, y^i, y^i_\la)$ of $J^1Y$
are compatible with these fibrations:
\[
x^\la \to {x'}^\la(x^\m),\qquad y^i \to {y'}^i(x^\m,y^j), \qquad
{y'}^i_\la = (\frac{\dr {y'}^i}{\dr y^j}y_\m^j +
\frac{\dr {y'}^i}{\dr x^\m})\frac{\dr x^\m}{\dr {x'}^\la}.
\]
There is the  canonical bundle monomorphism (the contact map)
\[
\la:J^1Y\op\to_Y
T^*X \op\otimes_Y TY, \qquad \la=dx^\la \otimes
(\dr_\la + y^i_\la \dr_i).
\]

Let  $\Phi$ be a fibred morphism  of $Y\to X$ to
$Y'\to X$ over a diffeomorphism of $X$. Its jet prolongation reads
\[
J^1\Phi : J^1Y\to  J^1Y', \qquad {y'}^i_\mu\circ
J^1\Phi=(\dr_\la\Phi^i+\dr_j\Phi^iy^j_\la)\frac{\dr x^\la}{\dr {x'}^\mu}.
\]
The jet prolongation of a section $s$ of $Y\to X$
is the section $y_\la^i\circ J^1s = \dr_\la s^i$ of $J^1Y\to X$.

The repeated jet manifold
$J^1J^1Y$, by definition, is the first order jet manifold of
$J^1Y\to X$. It is provided with the adapted coordinates
$(x^\la ,y^i,y^i_\la ,y_{(\m)}^i,y^i_{\la\m})$.
Its subbundle $ \wh J^2Y$ with $y^i_{(\la)}= y^i_\la$ is called the
sesquiholonomic jet manifold.
The second order jet manifold $J^2Y$ of $Y$ is the subbundle
of $\wh J^2Y$ with $ y^i_{\la\m}= y^i_{\m\la}.$

Given a fibred  manifold  $Y\to X$, a jet field
$\G$ on $Y$ is defined to be a  section
of the jet bundle $J^1Y\to Y$. A global jet field is a connection on $Y$.
By means of the contact map $\la$, every connection $\G$ on
$Y$ can be represented by the tangent-valued form
$\la\circ\G$ on $Y$. For the sake of simplicity, we  denote this
form by the same symbol
\[
\G =dx^\la\otimes(\dr_\la +\G^i_\la\dr_i).
\]

The Legendre manifold  $\Pi$ (\ref{00}) of a fibred manifold $Y$ is the
composite  manifold
\[
\pi_{\Pi X}=\pi\circ\pi_{\Pi Y}:\Pi\to Y\to X
\]
endowed with the fibred coordinates  $( x^\la ,y^i,p^\la_i)$:
\[
{p'}^\la_i = J \frac{\dr y^j}{\dr{y'}^i} \frac{\dr
{x'}^\la}{\dr x^\m}p^\m_j, \qquad J^{-1}=\det (\frac{\dr {x'}^\la}{\dr
x^\m}) .
\]
By $J^1\Pi$ is meant the first order jet manifold of  $\Pi\to X$.
It is provided with the adapted fibred coordinates
$
( x^\la ,y^i,p^\la_i,y^i_{(\m)},p^\la_{i\m}) .
$

By a momentum morphism, we call a fibred morphism
\[
\Phi: \Pi\op\to_Y J^1Y, \qquad ( x^\la ,y^i,y^i_\la)\circ\Phi= ( x^\la
,y^i,\Phi^i_\la(p)).
 \]
Given a momentum morphism $\Phi$, its composition with the
contact map $\la$
is represented by the  horizontal pullback-valued 1-form
\begin{equation}
\Phi =dx^\la\otimes(\dr_\la +\Phi^i_\la (p)\dr_i)\label{2.7}
\end{equation}
on $\Pi\to X$. For instance, let $\G$ be a connection on $Y\to X$. Then,
$ \wh\G=\G\circ\pi_{\Pi Y}$ is a momentum morphism. Conversely,
every momentum morphism $\Phi$ of
the Legendre manifold $\Pi$ of $Y$ defines
the associated connection $ \G_\Phi =\Phi\circ\wh 0$
on $Y\to X$ where $\wh 0$ is the global zero section of the
Legendre bundle $\Pi\to Y$.

\section{Lagrangian formalism}

Given a Lagrangian density  $L$, one can construct the exterior form
\begin{equation}
\La_L=(y^i_{(\la)}-y^i_\la)d\pi^\la_i\w\om +
(\dr_i-\wh\dr_\la\dr^\la_i)\cL dy^i\w\om,\label{304}
\end{equation}
\[
\la=dx^\la\otimes\wh\dr_\la, \qquad
\wh\dr_\la =\dr_\la +y^i_{(\la)}\dr_i+y^i_{\m\la}\dr^\m_i,
\]
on the repeated jet manifold $J^1J^1Y$.
Its restriction to the second order jet manifold $J^2Y$  reproduces
the familiar variational Euler-Lagrange operator
\begin{equation}
\cE_L= (\dr_i-\wh\dr_\la\dr^\la_i)\cL dy^i\w\om ,
\qquad \wh\dr_\la =\dr_\la +y^i_\la\dr_i+y^i_{\m\la}\dr^\m_i.
\label{305}
\end{equation}
The restriction of the form (\ref{304}) to the sesquiholonomic jet manifold
$\wh J^2Y$ of $Y$ defines the sesquiholonomic extension
\begin{equation}
\cE'_L:\wh J^2Y\to\op\w^{n+1}T^*Y \label{2.26}
\end{equation}
 of the Euler-Lagrange operator (\ref{305}).
 It has the form
(\ref{305}) with nonsymmetric coordinates $y^i_{\m\la}$.

 Let $\ol s$ be a section of the fibred jet manifold $J^1Y\to X$
such that its first order jet prolongation $J^1\ol s$ takes its values into
$\Ker\cE'_L$. Then, it satisfies the system of first order
Euler-Lagrange equations
 \begin{equation}
\dr_\la\ol s^i=\ol s^i_\la, \qquad
 \dr_i\cL-(\dr_\la+\ol s^j_\la\dr_j
+\dr_\la\overline s^j_\m\dr^\m_j)\dr^\la_i\cL=0. \label{306}
\end{equation}
They are equivalent to the  familiar second order
Euler-Lagrange equations
\begin{equation}
 \dr_i\cL-(\dr_\la+\dr_\la s^j\dr_j
+\dr_\la\dr_\mu s^j\dr^\m_j)\dr^\la_i\cL=0  \label{2.29}
 \end{equation}
for sections $s$ of $Y\to X$. We have $\ol s=J^1s$.

\section{Multimomentum  Hamiltonian formalism}

Let $\Pi$ be the Legendre manifold (\ref{00}) provided with the generalized
Liouville form $\theta$ (\ref{2.4}) and the multisymplectic form $\Om$
(\ref{406}).

We say that a connection
\[
\g =dx^\la\otimes(\dr_\la +\g^i_{(\la)}\dr_i
+\g^\m_{i\la}\dr^i_\m)
\]
on the Legendre manifold $\Pi\to X$ is a Hamiltonian connection
if the exterior form
\[
\g\rfloor\Om =dp^\la_i\w dy^i\w\om_\la
+\g^\la_{i\la}dy^i\w\om -\g^i_{(\la)} dp^\la_i\w\om
\]
is closed.

Hamiltonian connections
constitute an affine subspace of connections on
$\Pi\to X$. The following construction  shows that this subspace is
not empty.

Every connection $\G$ on  $Y\to X$
is lifted to the connection
\[
 \g=\wt\G =dx^\la\otimes[\dr_\la +\G^i_\la (y)\dr_i	 +
(-\dr_j\G^i_\la (y)  p^\m_i-K^\m{}_{\nu\la}(x) p^\nu_j+K^\al{}_{\al\la}(x)
p^\m_j)\dr^j_\m]
 \]
on $\Pi\to X$ where $K$ is a linear
symmetric connection on the bundles $TX$ and $T^*X$.
We have the equality
\begin{equation}
\wt\G\rfloor\Om =d(\wh\G\rfloor\th) \label{412}
\end{equation}
which shows that $\wt\G$ is a Hamiltonian connection.

An exterior $n$-form $H$ on the
Legendre manifold $\Pi$ is called a Hamiltonian form if
there exists a Hamiltonian connection for $H$ satisfying the equation
$\g\rfloor\Om  =dH.$

Note that Hamiltonian forms throughout are considered modulo
closed forms since closed
forms do not make any contribution into the Hamilton equations.

It follows that Hamiltonian  forms constitute an affine space
modelled on a linear space of the exterior horizontal densities
$\wt H=\wt{\cH}\om$ on $\Pi\to X$. A glance
at the equality (\ref{412}) shows that this affine space is not empty.
Given a connection $\G$ on a  $Y\to X$, its lift $\wt\G$ on
$\Pi\to X$ is a Hamiltonian connection for the Hamiltonian form
\begin{equation}
 H_\G=\wh\G\rfloor\th =p^\la_i dy^i\w\om_\la -p^\la_i\G^i_\la (y)\om.
\label{3.6}
\end{equation}
It follows that every Hamiltonian form on the Legendre
manifold $\Pi$ can be given by the expression (\ref{017}).

Moreover, a Hamiltonian form has the canonical splitting (\ref{017})
as follows. Every Hamiltonian form $H$
defines the associated momentum morphism
\[
\wh H:\Pi\to J^1Y, \qquad y_\la^i\circ\wh H=\dr^i_\la\cH,
\]
and the associated connection
$\G_H =\wh H\circ\wh 0$
on $Y\to X$. As a consequence, we have the canonical splitting
\begin{equation}
H=H_{\G_H}-\wt H.\label{3.8}
\end{equation}

The Hamilton operator $\cE_H$ for a Hamiltonian form $H$
is defined to be the first order differential operator
\begin{equation}
\cE_H=dH-\wh\Om=[(y^i_{(\la)}-\dr^i_\la\cH) dp^\la_i
-(p^\la_{i\la}+\dr_i\cH) dy^i]\w\om \label{3.9}
\end{equation}
where $\wh\Om$
is the pullback of the multisymplectic form $\Om$ onto $J^1\Pi$.

For any connection $\g$ on the Legendre manifold $\Pi$, we have
\[
\cE_H\circ\g =dH-\g\rfloor\Om.
\]
It follows that  $\g$  is a Hamiltonian jet field for a
Hamiltonian form $H$ if and only if it takes its values into
$\Ker\cE_H$, that is, satisfies  the algebraic Hamilton equations
\begin{equation}
\g^i_\la =\dr^i_\la\cH, \qquad
\g^\la_{i\la}=-\dr_i\cH. \label{3.10}
\end{equation}
Let $r$ be a section of the fibred Legendre manifold $\Pi\to X$ such that
its jet prolongation $J^1r$ takes its values into $\Ker\cE_H$.
Then, the Hamilton equations (\ref{3.10}) are brought to the first
order differential Hamilton equations (\ref{3.11a}) and (\ref{3.11b}).

Given a fibred manifold $Y\to X$, let $L$ be a first
order Lagrangian density. We shall say that a  Hamiltonian form
$H$ (\ref{017}) is associated with a Lagrangian
density $L$ if $H$ satisfies the relations
\bea
&&p^\la_j=\dr^\la_j\cL (x^\mu,y^i,\dr^i_\mu\cH),  \label{2.30a}, \\
&& \cH-p^\la_j\dr^j_\la\cH=\cL (x^\mu,y^i,\dr^i_\mu\cH ). \label{2.30b}
\eea
In the terminology of constraint theory, we call $Q=\wh L( J^1Y)$ the
constraint space.

If a Lagrangian density $L$ is regular,
there always exists the unique Hamiltonian form
associated with $L$, otherwise in general case.
In particular, all  Hamiltonian forms $H_\G$ (\ref{3.6}) are
associated with the Lagrangian density $ L=0.$

Contemporary field theories are almost never regular.
We shall restrict
our consideration to semiregular Lagrangian densities
$L$ when the preimage $\wh L^{-1}(p)$ of any point $p\in Q$ is a
connected submanifold of  $J^1Y$.
This notion of degeneracy seems most appropriate. Lagrangian densities
of fields are almost always semiregular.
In this case, one can get the workable relations between
Lagrangian and multimomentum Hamiltonian formalisms.

(i) A Hamiltonian form associated
with a semiregular Lagrangian density $L$ meets the condition
\[
\pi^\la_iy^i_\la-\cL=\cH(x^\m,y^i,\pi^\la_i).
\]

(ii) Let $H$ be a Hamiltonian form
associated with a semiregular Lagrangian density $L$. The Hamilton
operator $\cE_H$ for $H$ satisfies the relation
\[
\La_L=\cE_H\circ J^1\wh L.
\]

(iii) Let a  section $r$ of $\Pi\to X$
be a solution of the Hamilton equations (\ref{3.11a}) and (\ref{3.11b})
 for a Hamiltonian form $H$ associated with a semiregular Lagrangian
density $L$. If $r$ lives on the constraint space $Q$, the section
$\ol s=\wh H\circ r$ of $J^1Y\to X$ satisfies the first
order Euler-Lagrange equations (\ref{306}).
Conversely, given a semiregular Lagrangian density $L$, let
$\ol s$ be a solution of the
first order Euler-Lagrange equations (\ref{306}).
Let $H$ be a Hamiltonian form associated with $L$ so that
\begin{equation}
\wh H\circ \wh L\circ \ol s=\ol s.\label{2.36}
\end{equation}
Then, the section $r=\wh L\circ \ol s$ of $\Pi\to X$ is a solution of the
Hamilton equations (\ref{3.11a}) and (\ref{3.11b}) for $H$. It lives on the
constraint space $Q$.
Moreover, for every sections $\ol s$ and $r$ satisfying the
above-mentioned conditions, we have the relations
 \[
\ol s=J^1s, \qquad  s= \pi_{\Pi Y}\circ r
\]
where $s$ is a solution of the second order Euler-Lagrange equations
(\ref{2.29}).

We shall say that a family of Hamiltonian forms $H$
associated with a semiregular Lagrangian density $L$ is
complete if, for each solution $\ol s$ of the first order Euler-Lagrange
equations  (\ref{306}), there exists a
solution $r$ of the Hamilton equations
(\ref{3.11a}) and (\ref{3.11b}) for
some  Hamiltonian form $H$ from this family so that
\[
r=\wh L\circ\ol s,\qquad  \ol s =\wh H\circ r, \qquad
\ol s= J^1(\pi_{\Pi Y}\circ r).
\]
In virtue of assertion (iii), such a complete family
exists if and only if, for each solution $\ol s$ of the Euler-Lagrange
equations for $L$, there exists a  Hamiltonian form $H$ from this
family so that the condition (\ref{2.36}) holds.

In field models where Lagrangian densities are quadratic or affine in
velocities, there always exist complete families of associated
Hamiltonian forms.

\section{Hamiltonian gauge theory}

In the rest of the article, the manifold $X$ is assumed to be
oriented. It is
provided with a nondegenerate fibre metric $g_{\m\nu}$
in the tangent bundle of $X$. We denote $g=\det(g_{\m\nu}).$

Let $P\to X$ be a principal bundle with a structure Lie group $G$
wich acts on $P$ on the right by the law
\[
r_g: P\to Pg, \qquad g\in G.
\]
A principal connection is defined to be a
 $G$-equivariant global jet field $A$ on $P$:
 \[
A\circ r_g=J^1r_g\circ A, \qquad g\in G.
\]

There is the 1:1 correspondence between the principal connections $A$ on
$P$  and the global sections of the bundle $C=J^1P/G$.
It is the affine bundle modelled on the vector bundle
\[
\ol C =T^*X \otimes V^GP, \qquad  V^GP=VP/G.
\]

Given a bundle atlas $\Psi^P$ of $P$, the bundle $C$
is provided with  the fibred coordinates $(x^\mu,k^m_\mu)$ so that
\[
(k^m_\mu\circ A)(x)=A^m_\mu(x)
\]
are coefficients of the local connection 1-form of a principal connection
$A$ with respect to the atlas $\Psi^P$.

The first order jet manifold $J^1C$ of the bundle $C$ is
provided with the adapted coordinates $
(x^\mu, k^m_\mu, k^m_{\mu\lambda}).$
There exists the canonical splitting
\begin{equation}
J^1C=C_+\op\oplus_C C_-=(J^2P/G)\op\oplus_C
(\op\w^2 T^*X\op\otimes_C V^GP), \label{N31}
\end{equation}
\[
 k^m_{\mu\la}=\frac12(
k^m_{\mu\la}+k^m_{\la\mu} +c^m_{nl}k^n_\la k^l_\mu)
+\frac12( k^m_{\mu\la}-k^m_{\la\mu} -c^m_{nl}k^n_\la k^l_\mu),
\]
over $C$.
There are the corresponding canonical surjections:

(i) ${\cal S}: J^1 C\to C_+.$

(ii) $ \cF: J^1 C\to C_-$ where
\[
\cF=\frac{1}{2}\cF^m_{\la\m}dx^\la\w dx^\m\otimes e_m,
\qquad \cF^m_{\la\mu}=
k^m_{\mu\la}-k^m_{\la\mu} -c^m_{nl}k^n_\la k^l_\mu,
\]

The Legendre manifold of the bundle $C$ of principal connections reads
\[
\Pi=\op\w^n T^*X\otimes TX\op\otimes_C [C\times\ol C]^*.
\]
It is provided with the fibred coordinates
$(x^\mu,k^m_\mu,p^{\mu\la}_m) $
and has the canonical splitting
\[
\Pi=\Pi_+\op\oplus_C\Pi_-,
\]
\[
(p^{\mu\la}_m)=(p^{(\mu\la)}_m=\frac{1}{2}[p^{\mu\la}_m+
p^{\la\mu}_m]) +(p^{[\mu\la]}_m=\frac{1}{2}[p^{\mu\la}_m-p^{\la\mu}_m]).
\]

On the configuration space (\ref{N31}),
the conventional Yang-Mills Lagrangian density $L_{YM}$
is given by the expression
\begin{equation}
L_{YM}=\frac{1}{4\ve^2}a^G_{mn}g^{\la\mu}g^{\bt\nu}\cF^m_{\la
\beta}\cF^n_{\mu\nu}\sqrt{\mid g\mid}\,\om \label{5.1}
\end{equation}
where  $a^G$ is a nondegenerate $G$-invariant metric
in the Lie algebra of $G$. It is almost regular and semiregular.
The Legendre morphism
associated with the Lagrangian density (\ref{5.1}) takes the form
\bea
&&p^{(\mu\la)}_m\circ\wh L_{YM}=0, \label{5.2a}\\
&&p^{[\mu\la]}_m\circ\wh
L_{YM}=\ve^{-2}a^G_{mn}g^{\la\al}g^{\mu\bt}
\cF^n_{\al\bt}\sqrt{\mid g\mid}. \label{5.2b}
\eea

Let us consider
connections on the bundle $C$ of principal connections which
take their values into $\Ker\wh L_{YM}$:
\begin{equation}
S:C\to C_+, \qquad
S^m_{\m\la}-S^m_{\la\m}-c^m_{nl}k^n_\la k^l_\m=0. \label{69}
\end{equation}
For all these connections, the
Hamiltonian forms
\ben
&&H=p^{\mu\la}_mdk^m_\mu\w\om_\la-
p^{\mu\la}_mS^m_{\mu\la}\om-\wt{\cH}_{YM}\om, \label{5.3}\\
&&\wt{\cH}_{YM}= \frac{\ve^2}{4}a^{mn}_Gg_{\mu\nu}
g_{\la\bt} p^{[\mu\la]}_m p^{[\nu\bt]}_n\mid g\mid ^{-1/2},
\nonumber
\een
are associated with the Lagrangian density $L_{YM}$ and constitute the
complete family.
Moreover, we can minimize this complete family if we restrict our
consideration to connections (\ref{69}) of the following type.
Given a symmetric linear connection $K$
on the cotangent bundle $T^*X$ of $X$,  every principal connection $B$ on
$P$ is lifted to the connection $S_B$ (\ref{69}) such that
\[
S_B\circ B={\cal S}\circ J^1B,
\]
\begin{equation}
S_B{}^m_{\mu\la}=\frac{1}{2} [c^m_{nl}k^n_\la
k^l_\mu  +\dr_\mu B^m_\la+\dr_\la B^m_\mu -c^m_{nl}
(k^n_\mu B^l_\la+k^n_\la B^l_\mu)] -
K^\beta{}_{\mu\la}(B^m_\beta-k^m_\beta). \label{3.7}
\end {equation}
We denote the  Hamiltonian form (\ref{5.3}) for the
connections $S_B$ (\ref{3.7}) by $H_B$.

The corresponding Hamilton equations for sections $r$ of $\Pi\to X$ read
\ben
&&\dr_\la p^{\mu\la}_m=-c^n_{lm}k^l_\nu
p^{[\mu\nu]}_n+c^n_{ml}B^l_\nu p^{(\mu\nu)}_n
-K^\mu{}_{\la\nu}p^{(\la\nu)}_m, \label{5.5} \\
&&\dr_\la k^m_\mu+ \dr_\mu
k^m_\la=2S_B{}^m_{(\mu\la)}\label{5.6}
\een
plus Eqs.(\ref{5.2b}).
The
equations (\ref{5.2b}) and (\ref{5.5}) restricted to the constraint space
(\ref{5.2a}) are the familiar
Yang-Mills equations for $A=\p_{\Pi C}\circ r.$
Different Hamiltonian forms $H_B$ lead to different Eqs.(\ref{5.6}) which
play the role of the gauge-type condition.

\section{Hamiltonian Dirac equations}

As a test case let us consider Dirac fermion fields in the
presence of a background tetrad field $h$. Recall that they are represented
by global sections of the spinor bundle $S_h$ associated
with the $L_s$-lift of the reduced Lorentz subbundle $L^hX$ of the
linear frame bundle $LX$ \cite{sard10}. Their Lagrangian density
is defined on the configuration space $J^1(S_h\oplus S^*_h)$ provided
with the adapted coordinates
\[
( x^\m, y^A,y^+_A, y^A_\mu, y^+_{A\mu}).
\]
It reads
\ben
&&L_D=\{\frac{i}2[ y^+_A(\g^0\g^\m)^A{}_B( y^B_\m -A^B{}_{C\m}y^C)
 -( y^+_{A\m}-A^+{}^C{}_{A\m}y^+_C)(\g^0\g^\m)^A{}_By^B]\nonumber\\
&&\qquad -my^+_A(\g^0)^A{}_By^B\}h^{-1}\om,\label{5.60}
\een
\[
 \g^\m =h^\m_a( x)\g^a, \qquad h=\det (h^\mu_a),
\]
where
\[
A^A{}_{B\mu}=\frac12A^{ab}{}_\mu (x)I_{ab}{}^A{}_B
\]
is a principal connection on the principal spinor bundle $P_h$.

The Legendre bundle $\Pi_h$
over the spinor bundle $S_h\oplus S_h^*$ is provided
with the canonical coordinates
\[
( x^\m ,y^A,y^+_A,p^\m_A,p^{\m A}_+).
\]
Relative to these coordinates, the Legendre morphism associated
with the Lagrangian density (\ref{5.60}) is written
\ben
&&p^\m_A=\p^\m_A=\frac{i}2y^+_B(\g^0\g^\m)^B{}_Ah^{-1},\nonumber\\
&&p^{\m A}_+=\p^{\m A}_+=-\frac{i}2(\g^0\g^\m)^A{}_By^Bh^{-1}.
\label{5.61}
\een
It defines the constraint subspace of the Legendre bundle $\Pi_h$. Given a
soldering form
\[
S=S^A{}_{B\m}(x)y^Bdx^\m\otimes\dr_A
\]
 on the bundle $S_h$,
let us consider the connection $A+S$ on $S_h$. The corresponding
Hamiltonian forms associated with the Lagrangian
density (\ref{5.60}) read
\ben
&&H_S=( p^\m_A dy^A+p^{\m A}_+dy^+_A)\w\om_\m-\cH_S\om , \label{N54}\\
&&\cH_S=p^\m_AA^A{}_{B\m}y^B+y^+_BA^+{}^B{}_{A\m}p^{\m
A}_++my^+_A(\g^0)^A{}_By^Bh^{-1}+\nonumber\\
&&\qquad ( p^\m_A-\p^\m_A) S^A{}_{B\m} y^B+y^+_BS^+{}^B{}_{A\m}( p^{\m
A}_+-\p^{\m A}_+).\nonumber
\een

The corresponding Hamilton equations for a section $r$ of the fibred
Legendre manifold $\Pi\to X$
take the form
\bea
&&\dr_\m y^+_A=y^+_B( A^+{}^B{}_{A\m}+S^+{}^B{}_{A\m}) ,\label{5.62a}\\
&&\dr_\m p^\m_A=-p^\m_BA^B{}_{A\m}-( p^\m_B-\p^\m_B) S^B{}_{
A\m}- \nonumber \\
&& \qquad ( my^+_B(\g^0)^B{}_A
+\frac{i}{2}y^+_BS^+{}^B{}_{C\m}(\g^0\g^\m)^C{}_A)
h^{-1}\label{5.62b}
 \eea
plus the equations for the components $y^A$ and $p^{\m A}_+$.
The equation (\ref{5.62a})
and the similar equation for $y^A$ imply that $y$
is an integral
section for the connection $A+S$ on the spinor bundle $S_h$. It
follows that the Hamiltonian forms (\ref{N54})
constitute the complete family.

On the constraint
space (\ref{5.61}), Eq.(\ref{5.62b}) reads
\begin{equation}
\dr_\m\pi^\m_A=-\pi^\m_BA^B{}_{A\m}-(
my^+_B(\g^0)^B{}_A+ \frac{i}2y^+_BS^+{}^B{}_{C\m}(\g^0\g^\m)^C{}_A)
h^{-1}.\label{5.63}
\end{equation}
Substituting Eq.(\ref{5.62a}) into Eq.(\ref{5.63}), we
obtain the familiar Dirac equation in the presence of a tetrad
gravitational field $h$.

\section{Hamiltonian systems on composite manifolds}

In this Section, by $Y$ throughout is meant a composite manifold
\begin{equation}
 \pi:=\pi_{\Si X}\circ\pi_{Y\Si}:Y\to \Si\to X \label{1.34'}
\end{equation}
provided with the fibred coordinates $(x^\la,\si^m,y^i)$
where $(x^\m,\si^m)$ are fibred  coordinates  of the fibred manifold
$\Si\to X$. We further suppose that the fibred manifold
\[
Y_\Si :=Y\to\Si
\]
is a bundle.

Given the composite  manifold (\ref{1.34'}), let $J^1\Si$,
$J^1Y_\Si$ and $J^1Y$ be the first order jet manifolds of
$\Si\to X$, $Y\to \Si$ and $Y\to X$ respectively which are endowed
with the corresponding adapted coordinates:
\be
 &&( x^\la ,\si^m, \si^m_\la),\\
 &&( x^\la ,\si^m, y^i, \wt y^i_\la, y^i_m),\\
 &&( x^\la ,\si^m, y^i, \si^m_\la ,y^i_\la) .
\ee

Recall the following assertions \cite{sard,sard10}.

(i) Given a global section $h$ of the fibred manifold $\Sigma\to X$, the
restriction
\begin{equation}
Y_h=h^*Y_\Si \label{S10}
\end{equation}
of the bundle $Y_\Si$ to $h(X)$ is a fibred imbedded submanifold
 of the fibred manifold $Y\to X$.

(ii) There is the 1:1 correspondence between the
global sections $s_h$ of  $Y_h$ and the global sections of
the composite manifold $Y$ which cover the section $h$.

(iii) There exists the canonical surjection
\ben
&&\rho : J^1\Si\op\times_\Si J^1Y_\Si\to J^1Y,\label{1.38}\\
&&y^i_\la\circ\rho=y^i_m{\si}^m_{\la} +\wt y^i_{\la}.\nonumber
\een

Let
\begin{equation}
A_\Si=dx^\la\otimes(\dr_\la+\wt A^i_\la\dr_i)
+ d\si^m\otimes(\dr_m+A^i_m\dr_i) \label{Q1}
\end{equation}
be a connection on the bundle $Y\to \Si$ and
\[
 \G=dx^\la\otimes(\dr_\la + \G^m_\la\dr_m)
\]
a connection on the fibred manifold $\Sigma\to X$.
Building on the morphism (\ref{1.38}), one can construct the composite
connection
\begin{equation}
A=dx^\la\otimes[\dr_\la+\G^m_\la\dr_m +(A^i_m\G^m_\la + \wt A^i_\la)
\dr_i] \label{1.39}
\end{equation}
on the composite manifold $Y$. It possesses the following property.

Let $h$ be an integral section of the connection $\Gamma$ on
$\Si\to X$, that is,
\be
&&\G\circ h=J^1h,\\
&&\dr_\mu h^m=\G^m_\mu.
\ee
In this
case, the composite connection (\ref{1.39}) on $Y$ is reducible to the
connection
\begin{equation}
A_h=dx^\la\otimes[\dr_\la+(A^i_m\dr_\la h^m
+(\wt A\circ h)^i_\la)\dr_i] \label{1.42}
\end{equation}
on the fibred submanifold $Y_h$ (\ref{S10}) of $Y\to X$.

Given a composite manifold $Y$ (\ref{1.34'}),
every connection (\ref{Q1}) on the bundle $Y_\Si$ determines:
\begin{itemize}
\item the horizontal splitting
\begin{equation}
VY=VY_\Si\op\oplus_Y (Y\op\times_\Si V\Si),\label{46}
\end{equation}
\[
\dot y^i\dr_i + \dot\si^m\dr_m=
(\dot y^i -A^i_m\dot\si^m)\dr_i + \dot\si^m(\dr_m+A^i_m\dr_i),
\]
of the vertical tangent bundle $VY$ of $Y\to X$;
\item the dual horizontal splitting
\begin{equation}
V^*Y=V^*Y_\Si\op\oplus_Y (Y\op\times_\Si V^*\Si),\label{46'}
\end{equation}
\[
\dot y_i\ol dy^i + \dot\si_m\ol d\si^m=
\dot y_i(\ol dy^i -A^i_m\ol d\si^m) + (\dot\si_m +A^i_m\dot y_i)\ol d\si^m,
\]
of the vertical cotangent bundle $V^*Y$ of $Y\to X$.
\end{itemize}

It is readily observed that the splitting (\ref{46}) is uniquely
characterized by the form
\begin{equation}
\om\w A_\Si = \om\w d\si^m\otimes(\dr_m +A^i_m\dr_i). \label{227}
\end{equation}

Building on the horizontal splitting (\ref{46}), one can constract
the following first order differential operator on the composite manifold
$Y$:
\ben
&&\wt D={\rm pr}_1\circ D_A
: J^1Y\to T^*X\op\otimes_Y VY \to T^*X\op\otimes_Y VY_\Si,
\nonumber\\
 &&\wt D= dx^\la\otimes[y^i_\la-A^i_\la
-A^i_m(\si^m_\la-\G^m_\la)]\dr_i =\nonumber\\
&&\qquad  dx^\la\otimes(y^i_\la-\wt
A^i_\la -A^i_m\si^m_\la)\dr_i, \label{7.10}
 \een
where $D_A$ is the covariant differential relative to the
composite connection $A$ which is composition of $A_\Si$ and some
connection $\G$ on $\Si\to X$. We shall call $\wt D$ the vertical covariant
differential.

Let $h$ be an integral section of the connection $\G$ and $Y_h$ the portion
of $Y_\Si$ over $h(X)$.
It is readily observed that the vertical covariant differential
(\ref{7.10}) restricted to $J^1Y_h\subset J^1Y$
comes to the familiar covariant
differential for the connection $A_h$ (\ref{1.42}) on the
portion $Y_h\to X$ (\ref{S10}).

Thus, it is the vertical covariant differential (\ref{7.10}) that
we may utilize in order to construct a Lagrangian density
\begin{equation}
L: J^1Y\op\to^{\wt D} T^*X\op\otimes_YVY_\Si\to\op\w^nT^*X \label{229}
\end{equation}
for fields on a composite manifold. It should be noted that such a
Lagrangian density is never regular because of the constraint conditions
\[
\pi^\mu_iA^i_m=\pi^\mu_m.
\]
Therefore, the multimomentum Hamiltonian formalism must be applied.

The major feature of Hamiltonian systems on a composite manifold
$Y$ lies in the following.
The horizontal splitting (\ref{46'}) yields immediately
the corresponding splitting
of the Legendre bundle $\Pi$ over the composite manifold $Y$.
As a consequence,
the Hamilton equations (\ref{3.11a}) for sections $h$
of the fibred manifold $\Si\to X$ reduce to the gauge-type conditions
independent of momenta.
Thereby, these sections play the role of parameter fields.

Let $Y$ be a composite manifold (\ref{1.34'}). The Legendre bundle
$\Pi$ over $Y$ is endowed with the canonical coordinates
\[
(x^\la,\si^m,y^i,p^\la_m,p^\la_i).
\]
Let $A_\Si$ be a connection (\ref{Q1}) on the bundle $Y\to\Si$.
With a connection $A_\Si$, the splitting
\begin{equation}
\Pi=\op\w^nT^*X\op\otimes_YTX\op\otimes_Y
[V^*Y_\Si\op\oplus_Y (Y\op\times_\Si V^*\Si)]\label{230}
\end{equation}
of the Legendre bundle $\Pi$ is performed as an immediate consequence
of the splitting (\ref{46'}). We call this the horizontal splitting of
$\Pi$. Given the horizontal splitting (\ref{230}), the Legendre
bundle $\Pi$ is provided with the coordinates
 \ben
 &&\ol p^\la_i=p^\la_i, \nonumber\\
&& \ol p^\la_m = p^\la_m +A^i_mp^\la_i\label{231}
\een
which are compatible with this splitting.

Let $h$ be a global section of the fibred manifold $\Sigma\to X$.
Given the horizontal splitting (\ref{230}), the submanifold
\begin{equation}
\{\si=h(x), \ol p^\la_m=0\}\label{7.11}
\end{equation}
 of the Legendre bundle $\Pi$ over $Y$ is isomorphic to
the Legendre bundle $\Pi_h$ over the portion  $Y_h$ (\ref{S10}) of
 $Y_\Si$.

Let the composite manifold $Y$ be provided with the composite
connection (\ref{1.39}) determined by connections $A_\Si$ on $Y\to\Si$
and $\G$ on $\Si\to X$. Relative to the coordinates (\ref{231}) compatible
with the horizontal splitting (\ref{230}),
every Hamiltonian form on the Legendre bundle $\Pi$
over $Y$ can be given by the expression
\ben
&&H=(p^\la_idy^i+p^\la_md\si^m)\w\om_\la- \nonumber \\
&&[\ol p^\la_i\wt A^i_\la +\ol p^\la_m\G^m_\la
+\wt{\cH}(x^\mu, \si^m, y^i, \ol p^\mu_m, \ol p^\mu_i)]\om
\label{7.12}
\een
where
\[
\ol p^\la_i\wt A^i_\la +\ol p^\la_m\G^m_\la
= p^\la_iA^i_\la + p^\lambda_m\G^m_\la.
\]
 The corresponding Hamilton equations are
\bea
&&
\dr_\la p^\la_i=-p^\la_j[\dr_i\wt A^j_\la
+\dr_iA^j_m(\G^m_\la +\dr^m_\la\wt{\cH})]-\dr_i\wt{\cH},
\label{7.13a} \\
&&\dr_\la y^i=\wt A^i_\la +A^i_m(\G^m_\la
+\dr^m_\la\wt{\cH}) +\dr^i_\la\wt{\cH}, \label{7.13b} \\
&&\dr_\la p^\la_m= -p^\la_i[\dr_m\wt A^i_\la
+\dr_mA^i_n(\G^n_\la +\dr^n_\la\wt{\cH})] -\ol
p^\la_n\dr_m\G^m_\la -\dr_m\wt{\cH}, \label{7.13c} \\
&&\dr_\la\si^m=\G^m_\la +\dr^m_\la\wt{\cH} \label{7.13d}
\eea
and plus constraint conditions.

Let the Hamiltonian form
(\ref{7.12}) be associated with a Lagrangian density (\ref{229})
which contains the velocities $\si^m_\mu$ only inside the vertical
covariant differential (\ref{7.10}).
Then, the Hamiltonian density $\wt{\cH}\om$ appears independent of
the momenta $\ol p^\m_m$ and the Lagrangian constraints read
\begin{equation}
\ol p^\m_m=0. \label{232}
\end{equation}
In this case, Eq.(\ref{7.13d}) comes to the gauge-type condition
\begin{equation}
\dr_\la\si^m=\G^m_\la \label{237}
\end{equation}
independent of momenta.

 Let us consider now a Hamiltonian system in the presence of a background
parameter field $h(x)$.
Substituting Eq.(\ref{7.13d}) into Eqs.(\ref{7.13a})
- (\ref{7.13b}) and restricting them to the submanifold (\ref{7.11}), we
obtain the equations
\ben
&&
\dr_\la p^\la_i=-p^\la_j\dr_i[(\wt A\circ h)^j_\la
+A^j_m\dr_\la h^m]-\dr_i\wt{\cH},\nonumber \\
&&
\dr_\la y^i=(\wt A\circ h)^i_\la +A^i_m\dr_\la h^m
 +\dr^i_\la\wt{\cH}  \label{7.14}
\een
for sections of the Legendre manifold $\Pi_h\to X$  of the bundle $Y_h$
endowed with the connection (\ref{1.42}). Equations (\ref{7.14}) are the
Hamilton equations corresponding to the Hamiltonian form
\[
H_h=p^\la_idy^i\w\om_\la - [ p^\la_i A_h{}^i_\la
 +\wt{\cH}(x^\mu, h^m(x), y^i, p^\mu_i, \ol p^\mu_m=0)]\om
\]
on $\Pi_h$ which is induced by the Hamiltonian form (\ref{7.12}) on $\Pi$.

\section{Fermion-gravitation complex}

At first, we consider gravity without matter.

In the gauge gravitation theory, dynamic
gravitational variables are pairs of
tetrad gravitational fields $h$ and gauge gravitational potentials $A_h$
identified with principal connections on $P_h$. Following general procedure,
one can describe these  pairs $(h, A_h)$
by sections of the  bundle (\ref{N53}). The
corresponding  configuration space is the  jet manifold
$J^1C_L$ of $C_L$. The Legendre bundle
\begin{equation}
\Pi=\op\w^4 T^*X^4\op\otimes_{C_L} TX^4\op\otimes_{C_L} V^*C_L.
\label{5.54}
 \end{equation}
over $C_L$ plays the role of a phase space of the gauge
gravitation theory.

The bundle $C_L$ is endowed with the
local fibred coordinates
\[
(x^\mu,\si^\la_a,k^{ab}{}_\la=
-k^{ba}{}_\la, \si^\la_{a\mu})
\]
where
\[
(x^\mu,\si^\la_a, \si^\la_{a\mu})
\]
are coordinates of the jet bundle $J^1\Si$.
The jet manifold $J^1C_K$ of $C_K$ is provided with the corresponding
adapted coordinates
\[
(x^\mu,\si^\mu_a,k^{ab}{}_\la=
-k^{ba}{}_\la, \si^\mu_{a\la}=\si^\mu_{a(\la)},
k^{ab}{}_{\mu\la},\si^\mu_{a\la\nu}).
\]

The associated coordinates of the  Legendre manifold (\ref{5.54})
are
\[
(x^\mu, \si^\la_a, k^{ab}{}_\la, \si^\la_{a\nu},
 p^{a\mu}_\la, p_{ab}{}^{\la\mu}, p_\la^{a\nu\mu})
\]
where $(x^\mu, \si^\la_a, p^{a\mu}_\la) $ are
 coordinates of the Legendre manifold of the bundle $\Sigma$.

For the sake of simplicity, we here consider
the Hilbert-Einstein Lagrangian density of classical gravity
\begin{equation}
L_{HE}=-\frac{1}{2\kappa}\cF^{ab}{}_{\mu\la}\si^\mu_a
\si^\la_b\si^{-1}\om, \label{5.56}
 \end{equation}
\[
\cF^{ab}{}_{\mu\la}=k^{ab}{}_{\la\mu}-k^{ab}{}_{\mu\la}+
k^a{}_{c\mu} k^{cb}{}_\la-k^a{}_{c\la} k^{cb}{}_\mu,
\]
\[
 \si=\det(\si^\mu_a).
\]
The corresponding Legendre morphism $\wh L_{HE}$ is given by the coordinate
expressions
\bea
&&p_{ab}{}^{[\la\mu]}=\pi_{ab}{}^{[\la\mu]}=\frac{-1}{\kappa\si}
\si^{[\mu}_a\si^{\la]}_b. \label{5.57a}\\
&& p_{ab}{}^{(\la\mu)}=0, \qquad
p^{a\mu}_\la=0,\qquad
p^{a\nu\mu}_\la=0,  \label{5.57b}
\eea

We construct the complete family of multimomentum Hamiltonian forms
associated with the affine Lagrangian density (\ref{5.56}). Let
$K$ be a world connection associated with a principal connection $B$ on the
linear frame bundle $LX$. To minimize the complete family, we consider the
following connections on the bundle $C_K$:
 \be
 && \G^\la_{a\mu}=B^b{}_{a\mu}\si^\la_b -K^\la{}_{\nu\mu}\s^\nu_a,\\
&&\G^\la_{a\nu\mu}=\dr_\mu B^d{}_{a\nu}\si^\la_d
-\dr_\mu K^\la{}_{\bt\nu}\si^\bt_a \\
&& \qquad
+B^d{}_{a\mu}(\si^\la_{d\nu}-\G^\la_{d\nu})-K^\la{}_{\bt\mu}(\si^\bt_{a\nu}
-\G^\bt_{a\nu})+K^\bt{}_{\nu\mu}(\si^\la_{a\bt}-\G^\la_{a\bt})\\
&& \qquad +B^d{}_{a\nu}\G^\la_{d\mu}-K^\la{}_{\bt\nu}\G^\bt_{a\mu},\\
&&\G^{ab}{}_{\la\mu}=\frac12 [k^a{}_{c\la}k^{cb}{}_\mu-
k^a{}_{c\mu}k^{cb}{}_\la+\dr_\la B^{ab}{}_\mu+\dr_\mu B^{ab}{}_\la \\
&& \qquad -B^b{}_{c\mu}k^{ac}{}_\la -B^b{}_{c\la}k^{ac}{}_\mu
-B^a{}_{c\mu}k^{cb}{}_\la  -B^a{}_{c\la}k^{cb}{}_\mu]\\
&& \qquad
+K^\nu{}_{\la\mu}k^{ab}{}_\nu-K^\nu{}_{(\la\mu)}
B^{ab}{}_\nu -\frac12 R^{ab}{}_{\la\m},
\ee
where $R$ is the curvature of the connection $B$.

The complete family of multimomentum Hamiltonian forms associated
with the Lagrangian density (\ref{5.56}) consists of the forms given by the
coordinate expressions
 \be
 &&H_{HE}=(p_{ab}{}^{\la\mu}
dk^{ab}{}_\la+p^{a\mu}_\la d\si^\la_a +
p_\la^{a\nu\mu}d\si^\la_{a\nu})\w\om_\mu-\cH_{HE}\om,
\\
&&\cH_{HE}=(p_{ab}{}^{\la\mu} \G^{ab}{}_{\la\mu}+
p^{a\mu}_\la\G^\la_{a\mu}+p_\la^{a\nu\mu}
\G^\la_{a\nu\mu})+ \\
&& \qquad
\frac12 R^{ab}{}_{\la\mu}(p_{ab}{}^{[\la\mu]}-
\pi_{ab}{}^{\la\mu}).
\ee

The Hamilton equations corresponding to
such a multimomentum Hamiltonian form read
\bea
&& \cF^{ab}{}_{\mu\la}=R^{ab}{}_{\mu\la}, \label{5.58a}\\
&&\dr_{\mu}k^{ab}{}_{\la}+\dr_{\la}k^{ab}{}_{\mu}
=2\G^{ab}{}_{(\mu\la)}, \label{5.58b}\\
&& \dr_\mu\si^\la_a=\G^\la_{a\mu}, \label{5.58c}\\
&&\dr_\mu\si^\la_{a\nu}=\G^\la_{a\nu\mu}, \label{5.58d} \\
 &&\dr_\mu p_{ac}{}^{\la\mu}=-\frac{\dr\cH_{HE}}{\dr
k^{ac}{}_\la}, \label{5.58e}\\
&&\dr_\mu p^{a\mu}_\la=-\frac{\dr\cH_{HE}}{\dr\si^\la_a},
\label{5.58f}
\eea
plus the equations which are reduced to the trivial identities
on the constraint space (\ref{5.57a}). The equations (\ref{5.58a}) -
(\ref{5.58d}) make the sense of gauge-type
conditions. The equation (\ref{5.58d}) has the solution
 \[
\si^\la_{a\mu} =\dr_\nu\si^\la_a.
\]
The gauge-type condition (\ref{5.58b}) has the solution
\[
k(x)=B.
\]
It follows that the
forms $H_{HE}$ really constitute the complete family of multimomentum
Hamiltonian forms associated with the Hilbert-Enstein Lagrangian density
(\ref{5.56}).

On the constraint  space, Eqs.(\ref{5.58e}) and (\ref{5.58f})
are brought to the form
\bea
&&\dr_\mu\pi_{ac}{}^{\la\mu}=2k^b{}_{c\mu}\pi_{ab}{}^{\la\mu}+
\pi_{ac}{}^{\bt\g}\G^\la{}_{\bt\g},\label{5.59a}\\
 &&R^{cb}{}_{\bt\mu}\dr^a_\la\pi_{cb}{}^{\bt\mu}=0.\label{5.59b}
\eea
 The equation (\ref{5.59a}) shows that
$k(x)$ is the Levi-Civita connection for the tetrad field $h(x)$.
 Substitution of Eqs.(\ref{5.58a}) into Eqs.(\ref{5.59b}), leads
to the familiar Einstein equations.

Turn now to the fermion matter.
Given the $L_s$-principal lift $P_\Si$ of $LX_\Si$,
let us consider the composite spinor bundle
\begin{equation}
S:=\pi_{\Si X}\circ\pi_{S\Si}:(P_\Si\times V)/L_s\to\Si\to X^4 \label{L5}
\end{equation}
where
\[
S_\Si:=S\to\Si
\]
is the spinor bundle associated with the $L_s$ principal bundle $P_\Si$.
It is readily observed
that, given a global section $h$ of the Higgs bundle
$\Si\to X^4$, the restriction $S\to
\Si$ to $h(X^4)$ consists with the spinor bundle $S_h$
whose sections describe Dirac
fermion fields in the presence of the background tetrad field $h$.

Let us provide the principal bundle $LX$ with a holonomic atlas
$\{U_\xi,\psi^T_\xi\}$ and the principal bundles $P_\Si$ and $LX_\Si$
with associated atlases $\{U_\e, z^s_\e \}$ and
\[
\{U_\e, z_\e=r\circ z^s_\e\}
\]
respectively. Relative to these atlases, the composite spinor bundle
(\ref{L5}) is endowed
with the fibred coordinates
\[
(x^\la,\si_a^\mu, y^A)
\]
where $(x^\la,
\si_a^\mu)$ are fibred coordinates of the Higgs bundle $\Si\to X$
which is coordinatized by matrix components $\si^\mu_a$ of the
group elements
\[
GL_4\ni (\psi^T_\xi\circ z_\e)(\si): {\bf R}^4\to {\bf R}^4,\qquad
 \si\in U_\e,\qquad \pi_{\Si X}(\si)\in U_\xi.
\]
Given a section $h$ of $\Si\to X^4$, we have
\be
&&z^h_\xi (x)= (z_\e\circ h)(x),\\
&& (\si^\la_a\circ h)(x)= h^\la_a(x),\\
&& h(x)\in U_\e, \qquad x\in U_\xi,
\ee
where $h^\la_a(x)$ are the tetrad functions.

The jet manifolds $J^1\Si$, $J^1S_\Si$ and $J^1S$ of the bundles $\Si$,
$S_\Si$ and $S$ respectively are provided with the
adapted coordinates
\be
&&(x^\la,\si^\mu_a, \si^\mu_{a\la}),\\
&&(x^\la,\si^\mu_a, y^A,\wt y^A_\la, y^A{}^a_\mu),\\
&&(x^\la,\si^\mu_a, y^A,\si^\mu_{a\la}, y^A_\la).
\ee

Let us consider the bundle of Minkowski spaces
\[
(LX\times M)/L\to\Si
\]
associated with the $L$-principal bundle $LX_\Si$. It is isomorphic to
the pullback $\Si\times T^*X$ which we denote by the same symbol
$T^*X$. We have the bundle morphism
\begin{equation}
\g_\Si: T^*X\op\otimes_\Si S_\Si= (P_\Si\times (M\otimes V))/L_s\to
 (P_\Si\times\g(M\otimes V))/L_s=S_\Si, \label{L7}
\end{equation}
\[
\wh dx^\la=\g_\Si (dx^\la) =\si^\la_a\g^a,
\]
where $dx^\la$ is the basis for the fibre of $T^*X$ over $\si\in\Si$.
Owing to the canonical vertical splitting
\[
VS_\Si =S_\Si\op\times_\Si S_\Si,
\]
the morphism (\ref{L7}) implies the corresponding morphism
\begin{equation}
\g_\Si: T^*X\op\otimes_SVS_\Si\to VS_\Si. \label{L8}
\end{equation}

To construct a connection on the composite spinor bundle (\ref{L5}),
let us consider a connection on the composite bundle
\begin{equation}
LX\to \Si\to X^4.\label{01}
\end{equation}
Given a principal connection
\[
A_\Si = (\wt A^{ab}{}_\mu, A^{ab}{}^c_\mu)
\]
on $LX\to\Si$ and a connection $\G^\nu_{c\mu}$ on $\Si$, let
\[
A=(A^{ab}{}_\mu, \G^\nu_{c\mu}), \qquad
A^{ab}{}\mu=\wt A^{ab}{}_\mu +\G^\nu_{c\mu} A^{ab}{}^c_\nu,
\]
be the composite connection (\ref{1.39}) on (\ref{01}). We require
that, given a tetrad gravitational field $h$, its reduction (\ref{1.42})
\[
A_h{}^{ab}{}_\mu= \wt A^{ab}{}_\mu +\dr_\mu h^\nu_c A^{ab}{}^c_\nu
\]
consists with the Levi-Civita connection. Then, it is readily observed
that $A^{ab}{}_\mu$ must be given by the relation (\ref{M4}) and
\begin{equation}
\wt A^{ab}{}_\mu=\frac12 K^\nu{}_{\la\mu}\si^\la_c (\eta^{ca}\si^b_\nu
-\eta^{cb}\si^a_\nu ) \label{L10}
\end{equation}
where $K$ is some symmetric connection on $TX$. Then, the associated
connection on the spinor bundle $S\to\Si$ reads
\[
A_\Si=dx^\la\otimes (\dr_\la +\frac12\wt A^{ab}{}_\la
I_{ab}{}^B{}_Ay^A\dr_B) + d\si^\mu_c\otimes
(\dr^c_\mu+\frac12 A^{ab}{}^c_\mu I_{ab}{}^B{}_Ay^A\dr_B) .
\]
It determines the canonical horizontal
splitting (\ref{46}) of the vertical tangent bundle $VS_\Si$ given by
the form (\ref{227})
\[
\om\w\otimes [\dr^C_\mu +A^B{}^C_\mu\dr_B].
\]

The total configuration space of the fermion-gravitation complex is
the product
\[
J^1S\op\times_{J^1\Si}J^1C_L.
\]
On this configuration space, the Lagrangian density $L_{FG}$ of the
fermion-gravitation complex is the sum of the Hilbert-Einstein
Lagrangian density $L_{HE}$ (\ref{5.56}) and the modification $L_{D'}$
of the Lagrangian density (\ref{5.60}) of fermion fields:
\be
&&L_{D'}=\{\frac{i}2[ y^+_A(\g^0\g^\m)^A{}_B( y^B_\m -\frac12
(k^{ab}{}_\mu -A^{ab}{}^c_\nu(\si^\nu_{c\mu}-\G^\nu_{c\mu}))
I_{ab}{}^B{}_{C\m}y^C) -\\
&& \quad ( y^+_{A\m}-\frac12
(k^{ab}{}_\mu -A^{ab}{}^c_\nu(\si^\nu_{c\mu}-\G^\nu_{c\mu}))
I^+_{ab}{}^C{}_{A\m}y^+_C)(\g^0\g^\m)^A{}_By^B]
-my^+_A(\g^0)^A{}_By^B\}\si^{-1}\om
\ee
where
\[
\g^\mu=\si^\mu_a\g^a.
\]

The total phase space $\Pi$ of the fermion-gravitation complex is
coordinatized by
\[
(x^\la,\si^\mu_c,\si^\mu_{c\nu},y^A,y^+_A,k^{ab}{}_\mu, p^{c\la}_\mu,
p^{c\nu\la}_\mu,p^\la_A, p^{A\la}_+, p_{ab}{}^{\mu\la})
\]
and admits the corresponding splitting (\ref{230}). The Legendre morphism
associated with the Lagrangian density $L_{FG}$ defines the constraint
subspace of $\Pi$ given by the relations (\ref{5.61}), (\ref{5.57a}) and
conditions
\ben
&& p_{ab}{}^{(\la\mu)}=0,\nonumber\\
&& p^{c\nu\mu}_\la =0,\nonumber\\
&& \frac12 p^\mu_AA^{ab}{}^c_\nu I_{ab}{}^A{}_Cy^C +
\frac12 p_+^{A\mu} A^{ab}{}^c_\nu I^+_{ab}{}^C{}_Ay^+_C +p^{c\mu}_\nu=0.
\label{L12}
\een

Hamiltonian forms associated with the Lagrangian density $L_{FG}$ are
the sum of the Hamiltonian forms $H_{HE}$ and $H_S$ (\ref{N54}) where
\begin{equation}
A^A{}_{B\mu}=\frac12 k^{ab}{}_\mu I_{ab}{}^A{}_By^B. \label{L15}
\end{equation}
The corresponding Hamilton equations for spinor fields consist with
Eqs.(\ref{5.62a}) and (\ref{5.62b}) where $A$ is given by the
expression (\ref{L15}). The Hamilton equations (\ref{5.58a}) -
(\ref{5.58d}) remain true. The Hamilton equations (\ref{5.58e}) and
(\ref{5.58f}) contain additional matter sources. On the constraint
space
\[
p^{a\mu}_\la =0
\]
the modified equations (\ref{5.58f}) would come to the familiar Einstein
equations
\[
G^a_\mu +T^a_\mu=0
\]
where $T$ denotes the energy-momentum tensor of fermion fields,
otherwise on the constraint space (\ref{L12}). In that latter case,
we have
\begin{equation}
D_\la p^{c\la}_\mu=G^a_\mu +T^a_\mu \label{L16}
\end{equation}
where $D_\mu$ denotes the covariant derivative with respect to the
Levi-Civita connection which acts on the indices $^c_\mu$. Substitution
of (\ref{L12}) into (\ref{L16}) leads to the modified Einstein equations
for the total system of fermion fields and gravity:
\[
-\frac12 J^\la_{ab} D_\la A^{ab}{}^c_\mu=G^c_\mu +T^c_\mu
\]
where $J$ is the spin current of the fermion fields.

\end{document}